\documentclass[aps,prl,twocolumn,preprintnumbers,superscriptaddress,floatfix,nofootinbib,notitlepage,showkeys,showpacs]{revtex4-1}

\usepackage[utf8x]{inputenc}
\usepackage{titlesec}
\usepackage{graphicx}
\usepackage{hyperref}
\usepackage{latexsym}
\usepackage{amsmath}
\usepackage{amssymb}
\usepackage{bbm}
\usepackage{ulem}
\usepackage{pdfsync}
\usepackage{epsfig}
\usepackage{epstopdf}
\usepackage{color}
\usepackage{comment}
\usepackage{placeins}
\usepackage{slashed}
\usepackage{mathrsfs}

\newcommand{\mpl}{M_{P}}
\newcommand{\td}{{\rm d}}

\newcommand{\ie}{{\it i.e.}}
\newcommand{\eg}{{\it e.g.\,}}

\newcommand{\MHz}{{\rm MHz}}

\newcommand{\eV}{{\rm eV}}
\newcommand{\keV}{{\rm keV}}
\newcommand{\MeV}{{\rm MeV}}
\newcommand{\GeV}{{\rm GeV}}
\newcommand{\TeV}{{\rm TeV}}

\newcommand{\be}{\begin{equation}}
\newcommand{\ee}{\end{equation}}
\newcommand{\br}{\begin{eqnarray}}
\newcommand{\bea}{\begin{equation}\begin{aligned}}
\newcommand{\eea}{\end{aligned}\end{equation}} 
\newcommand{\er}{\end{eqnarray}}
\newcommand{\ba}{\begin{array}}
\newcommand{\ea}{\end{array}}
\newcommand{\bi}{\begin{itemize}}
\newcommand{\ei}{\end{itemize}}
\newcommand{\bn}{\begin{enumerate}}
\newcommand{\en}{\end{enumerate}}
\newcommand{\bc}{\begin{center}}
\newcommand{\ec}{\end{center}}
\newcommand{\Eq}[1]{Eq.~(\ref{#1})}

\newcommand{\gsim}{\lower.7ex\hbox{$\;\stackrel{\textstyle>}{\sim}\;$}}
\newcommand{\lsim}{\lower.7ex\hbox{$\;\stackrel{\textstyle<}{\sim}\;$}}

\begin{document}

\title{The EDGES 21 cm Anomaly and Properties of Dark Matter}

\author{Sean Fraser}
\affiliation{NICPB, R\"avala 10, 10143 Tallinn, Estonia}

\author{Andi Hektor}
\affiliation{NICPB, R\"avala 10, 10143 Tallinn, Estonia}

\author{Gert H\"utsi}
\affiliation{NICPB, R\"avala 10, 10143 Tallinn, Estonia}

\author{Kristjan Kannike}
\affiliation{NICPB, R\"avala 10, 10143 Tallinn, Estonia}

\author{Carlo Marzo}
\affiliation{NICPB, R\"avala 10, 10143 Tallinn, Estonia}

\author{Luca Marzola}
\affiliation{NICPB, R\"avala 10, 10143 Tallinn, Estonia}

\author{Christian Spethmann}
\affiliation{NICPB, R\"avala 10, 10143 Tallinn, Estonia}

\author{Antonio Racioppi}
\affiliation{NICPB, R\"avala 10, 10143 Tallinn, Estonia}

\author{Martti Raidal}
\affiliation{Theoretical Physics Department, CERN, Geneva, Switzerland}
\affiliation{NICPB, R\"avala 10, 10143 Tallinn, Estonia}

\author{Ville Vaskonen}
\affiliation{NICPB, R\"avala 10, 10143 Tallinn, Estonia}

\author{Hardi Veerm\"ae}
\affiliation{Theoretical Physics Department, CERN, Geneva, Switzerland}
\affiliation{NICPB, R\"avala 10, 10143 Tallinn, Estonia}

\begin{abstract}
The recently claimed anomaly in the measurement of the 21 cm hydrogen absorption signal by EDGES at $z\sim 17$, if cosmological, requires the existence of new physics. The possible attempts to resolve the anomaly rely on either {\it (i)} cooling the hydrogen gas via new dark matter-hydrogen interactions or  {\it (ii)} modifying the soft photon background beyond the standard CMB one, as possibly suggested also by the ARCADE~2 excess. We argue that solutions belonging to the first class are generally in tension with cosmological dark matter probes once simple dark sector models are considered. Therefore, we propose soft photon emission by light dark matter as a natural solution to the 21 cm anomaly, studying a few realizations of this scenario. We find that the signal singles out a photophilic dark matter candidate characterised by an enhanced collective decay mechanism, such as axion mini-clusters.
\end{abstract}

\maketitle


\section{Introduction}
\label{sec:Introduction}

The 21 cm signal of atomic hydrogen from the dark ages and cosmic dawn provides important information on the thermal and ionization history of the Universe. Recently, the low-band antenna of the Experiment to Detect the Global EoR Signature (EDGES) has reported an anomalously strong absorption in the measured 21 cm signal at redshifts in the range of $z\approx 13.2-27.4$~\cite{Bowman:2018aa}.
The anomaly could be a new signal of baryon-dark matter (DM) interaction~\cite{Tashiro:2014tsa,Munoz:2015bca},
as scattering processes may over-cool the hydrogen gas with respect to the standard expectations from the Cosmic Microwave Background (CMB) measurements.
Alternatively, the anomaly may be related to, or even be a consequence of~\cite{Feng:2018rje}, the excess in the radio background observed by ARCADE~2~\cite{2011ApJ...734....5F}. Most radical explanations advocate instead for a purely baryonic Universe~\cite{McGaugh:2018ysb} and, regardless of the actual mechanism behind the 21 cm anomaly, it is clear that the signal offers  new probes for the physics beyond the standard model of particle interactions.
In this paper we therefore analyze new physics scenarios that explain the EDGES 21~cm absorption anomaly, pointing out important complementary implications. 

On general grounds, the intensity of the observable 21~cm signal is proportional to~\cite{Bowman:2018aa}
\bea
I_{21}\propto 1- \frac{T_R(z)}{T_S(z)},
\label{signal}
\eea
where $T_S$ is the spin temperature of the gas and $T_R$ is temperature of radiation for a fiducial $z\sim 17$. The form of \Eq{signal} suggests that, 
if the EDGES result is indeed cosmological, there are {\it only two} potential ways for new physics to resolve the anomaly.

Firstly, the introduction of new interactions can lower the hydrogen spin temperature while keeping $T_R=T_{\rm CMB}$. This approach has been followed in recent papers~\cite{Barkana:2018aa,Fialkov:2018xre,Munoz:2018pzp,Berlin:2018sjs} which introduce a new baryon-DM velocity-dependent interaction~\cite{Barkana:2018aa,Fialkov:2018xre} or a milli-charged DM fraction~\cite{Munoz:2018pzp,Berlin:2018sjs}. As we argue below, addressing the 21 cm anomaly in this way forces to confront cosmological measurements that bound the scenario severely~\cite{Munoz:2017qpy,Gluscevic:2017ywp,Xu:2018efh}. 

Alternatively, it is possible to consider new physics scenarios that increase $T_R$ by emission of extra soft photons in the early Universe, addition to the standard CMB background. This explanation is also in line with the ARCADE~2 excess, requiring that only few percent of the ARCADE~2 photons must be of cosmological origin to explain also the EDGES anomaly. A first example of this approach was given in the recent paper~\cite{Ewall-Wice:2018bzf}, where the photon emission by accreting intermediate mass black holes is analysed. Light WIMP annihilations or decays that might be consistent with the ARCADE~2 excess~\cite{Fornengo:2011cn} also belong to this class of scenarios.

In the following, to be as general as possible, we first address the above mentioned solutions to the 21 cm anomaly and then propose new ones. In agreement with Ref.~\cite{Berlin:2018sjs}, we demonstrate that the proposed simple scenarios relying on extra gas cooling typically face generic and robust constraints. This motivates us to study scenarios where DM couples to 
photons producing a new soft radiation component with $T_R>T_{\rm CMB}$. In particular, we show that the scenarios that produce a hard photon background in addition to the soft counterpart are also severely constrained. In light of this, axions~\cite{Peccei:1977hh,Weinberg:1977ma}, axion-like particles (ALPs)~\cite{Ringwald:2012hr,Marsh:2015xka}, light oscillating spin-2 DM~\cite{Marzola:2017lbt} and light excited DM~\cite{Finkbeiner:2014sja} seem to be the DM candidates favoured by the 21 cm anomaly. Nevertheless, we find that even these scenarios need to be augmented by an enhanced collective decay mechanism, such as the decay via parametric resonance, in order to explain the signal.

\section{Hydrogen cooling from DM scattering }
\label{sec:xs}

We start by reviewing the possibility that hydrogen-DM elastic scatterings may cool the gas and improve the consistency with data~\cite{Barkana:2018aa,Fialkov:2018xre,Munoz:2018pzp}. To model the case, Ref.~\cite{Barkana:2018aa,Fialkov:2018xre} {\it assumed} a velocity-dependent elastic cross section of the form
\bea
\sigma(v)=\sigma_1 \left(  \frac{v}{1{\rm km\, s}^{-1}} \right)^{-n},
\label{sigma}
\eea 
where the reference value $\sigma_1>3.4\times 10^{-21}\,{\rm cm}^2=3.4~\times~10^3\,{\rm b}$ is taken at the velocity $v=1\,{\rm km\, s}^{-1}$, and
the value of $n$ reflects the velocity dependence of the scattering cross section. It is crucial to notice that dependences as strong as $n=4$ are needed in order to suppress the cross section at the recombination epoch and comply with the existing CMB and cosmological bounds~\cite{Gluscevic:2017ywp,Xu:2018efh}.
The choice $n=4$ in \Eq{sigma} can be motivated by the non-relativistic Rutherford cross section, i.e. by the QED of point-like particles.
However, only a tiny fraction  of the hydrogen, $x_{e} \approx 10^{-4}$, is in the ionized form at the epoch the 21 cm signal is generated.
Thus, the dominant fraction of the atomic hydrogen gas possess only dipole  interactions (see for example Ref.~\cite{0370-1328-71-6-301} for explicit computations) which, instead of the assumed $v^{-4}$ behaviour, predicts  scattering processes with $n=2$. The latter are firmly excluded by the CMB alone. We stress that this argument is completely general. 
The simplified approaches to the 21 cm anomaly seem then to face equally simple constraints, and more detailed scenarios must be formulated to study their actual consistency with data.

An example of more detailed scenario is presented by millicharged DM~\cite{Munoz:2018pzp,Berlin:2018sjs}.
In this case the momentum transfer cross-section is given by
\begin{equation}
  \bar{\sigma}_{t} = \frac{2 \pi \alpha^{2} \epsilon^{2} \xi}{\mu_{\chi,t}^{2} v^{4}},
  \label{eq:loeb:cross:section}
\end{equation}
where $\epsilon$ is the millicharge of the DM particle, $\mu_{\chi,t}$ is the DM-target reduced mass, $\alpha$ is the fine-structure constant, $v$ is the relative velocity between the two particles and $\xi$ is the Debye logarithm. The cross-section \eqref{eq:loeb:cross:section} arises from scattering with the ionized fraction $x_{e}(z=20) \approx 10^{-4}$ of hydrogen atoms.  With $\epsilon = 10^{-6}$ and $\mu_{\chi,t} \approx m_{e} = 0.5$~MeV, the cross section is about $\bar{\sigma}_{t} = 4 \times 10^{-12}~\mathrm{cm}^{-2}$ at $z \approx 20$. The fraction of milli-charged DM is taken to be $f_{\rm DM} \approx 0.1$. However, at the time of photon decoupling ($z \approx 1100$) we have $x_{e}(z=1100) \approx 1$, so the full cross-section at that time $\sigma = x_{e}(z=1100) \, f_{\mathrm{DM}} \, \bar{\sigma}_{t} \approx 5 \times 10^{-20}~\mathrm{cm}^{-2}$, while the CMB bound is $\sigma \approx 10^{-26}~\mathrm{cm}^{-2}$ for $m_{\chi} = 10$~MeV.

Our result agrees with the one in Ref.~\cite{Berlin:2018sjs}, where the authors state that the millicharged DM is allowed only in the small range $m_{\chi} \sim 10 - 80$~MeV, $\epsilon \sim 10^{-6} - 10^{-4}$ and $f_{\rm DM} \sim 0.003 - 0.02$.

 Although this type of arguments cannot exclude all possible solutions of hydrogen cooling based on new DM interactions, they demonstrate that these scenarios are severely constrained and that their assessment requires a careful study. For these reasons, we prefer to study here another class of solutions to the 21 cm anomaly, which relies on $T_R>T_{\rm CMB}$ in \Eq{signal}.

\section{Soft vs hard photon background} 
\label{sec:hard}

Most of the conventional DM candidates predict that DM processes inject charged cosmic rays as well as photons into the Universe.
Light, sub-GeV WIMPs represent the most well-known type of DM with such properties that has excited our community in last few years.
Sub-GeV dark matter in the keV to MeV range can in the future be probed by detectors based on Fermi-degenerate materials \cite{Hochberg:2015fth}, superconductivity \cite{Hochberg:2015pha}, superfluid helium \cite{Knapen:2016cue,Schutz:2016tid}, semiconductors \cite{Hochberg:2016sqx} or Dirac materials \cite{Hochberg:2017wce}. Superconducting detectors \cite{Hochberg:2016ajh} or detectors based on resonant dark matter absorption in molecules \cite{Arvanitaki:2017nhi} could bring the detectable mass range down to eV range.
Even the LHC monojet \cite{Aaboud:2017phn,Sirunyan:2017jix} and monophoton \cite{Aaboud:2017dor,Sirunyan:2017ewk} searches, which are typically insensitive for such a low mass range, may still constrain these scenarios significantly.

In our case, to explain the 21 cm anomaly, the WIMP mass must be precisely in the sub-GeV range in order for their annihilations or decays to be compatible with the charged cosmic ray~\cite{Cirelli:2010xx} and extra-galactic gamma-ray~\cite{Hutsi:2010ai,Finkbeiner:2010sm} backgrounds. 
These processes were regarded as compatible with the ARCADE~2 excess~\cite{Fornengo:2011cn}, although later measurements tended to disfavor this interpretation~\cite{2015MNRAS.447.2243V}. 
In fact, the predicted photon spectrum is generally broad and includes energetic photons in addition to the soft spectrum needed to explain the 21 cm anomaly, which make it necessary to address the constraints on the hard part of the spectrum. 

Similarly, if a fraction of the DM consist of primordial black holes~\cite{Carr:1974nx}, that is constrained to be below unity for most of the parameter 
space \cite{Carr:2017jsz},\footnote{Light wormholes or other horizonless objects can form all the DM~\cite{Raidal:2018eoo}. However, that light objects are expected to radiate hard spectrum, thus intermediate mass black holes are more appropriate to explain the soft photon excess~\cite{Ewall-Wice:2018bzf}.} 
the accretion of matter onto these systems may indeed explain the 21 cm anomaly~\cite{Ewall-Wice:2018bzf}. However, also in this case the photon spectrum is broad, leading to a hard radiation part that needs to be addressed.

As stated above, any mechanism which injects photons into the cosmic medium has to be such that it does not lead to an extra heating of the gas by the hard radiation. In fact, assuming that the heating of gas is negligible, reproducing the required 21 cm absorption feature requires the standard radiation field temperature $T_{\rm CMB}$ to be increased by a factor of $\sim 7/4$. In practise, extra photons are needed to provide $\sim 3/4 \, T_{\rm CMB}$. Thus, the amount of background soft photons must be doubled in the frequency range $(65-90)\times(1+z)$ MHz of the absorption feature, for $z \approx 17$. For example, if the injection spectrum has a spectral index $\alpha=1$, i.e.  $I_{\nu} \propto \nu^{-\alpha}$, the energy density of the extra photons in the $(65-90)\times 18$ MHz band corresponds to $\sim 3\times 10^{-6}$~eV/cm$^3$, while the CMB photons provide $\sim 4\times 10^{-6}~\mathrm{eV}/\mathrm{cm}^3$. The kinetic energy density $\rho_k$ of the gas at $z=17$, where the gas temperature is $T_g\sim 7$~K, can be estimated as $\rho_k\sim 1.5\times 10^{-6}$~eV/cm$^3$, and therefore is approximately two times lower than the energy density of the injected photons in the narrow energy band considered. As the most relevant absorption process for the soft photons injected after the recombination epoch is the free-free absorption, which turns out to provide quite a negligible opacity~\cite{Chluba:2015hma}, for these energies it is not necessary to worry about the gas heating. However, this is not the case if the energy injection occurs over much wider energy range, as is the case for black holes or annihilating/decaying WIMP DM. 

\begin{figure}[t]
\begin{center}
\includegraphics[width=.44\textwidth]{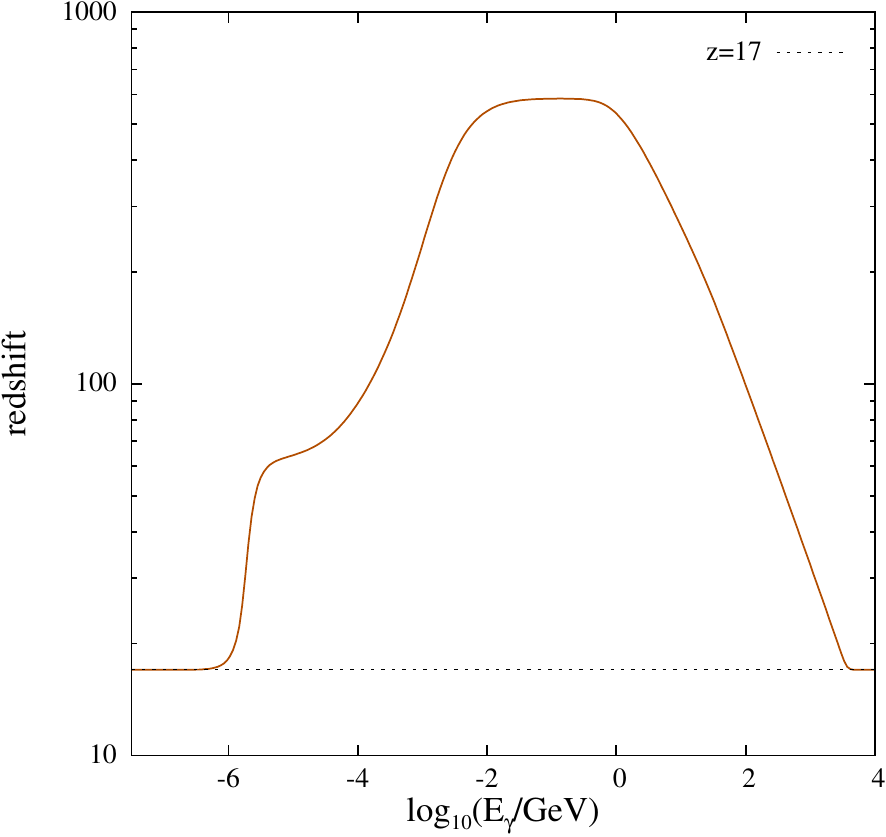}
\caption{Photosphere of the Universe (\ie, optical depth $\tau=1$) for a broad range of physical photon energies.  The observer is assumed to be located at redshift $z=17$. In the lowest energy section of the plot, before the first small plateau, the energy losses are dominated by photoionizations. Beyond that, up to the beginning of the large plateau, the Compton losses are dominating. At the large flat plateau region the energy loss is dominated by the pair production on matter. On the final falling part of the curves the energy losses are determined by photon-photon pair production. The calculations use results presented in~\cite{1989ApJ...344..551Z}.}
\label{fig:photosphere}
\end{center}
\end{figure}

In more detail, black holes have emission spectra with spectral indices $\alpha\sim 1$, \eg~\cite{1994ApJS...95....1E} over a broad range of energies. Thus, in this case, equal logarithmic energy bins provide equal energy input. It is clear that after normalizing the injection at 21 cm, a large fraction of the emitted energy falls into UV part of the spectrum, and gets absorbed due to a large photoionization opacity of the neutral medium. A significant fraction of the absorbed energy will then heats the gas~\cite{1985ApJ...298..268S}. In Fig.~\ref{fig:photosphere} we show the photosphere of the Universe, i.e. the redshift at which the optical depth reaches unity, for a broad range of physical photon energies. Indeed, one can see that the photons with energies less than $\sim 1$~keV get very efficiently absorbed.

\section{Photophilic Dark Matter} 
\label{sec:Results}

To explain the EDGES anomaly we then require a narrow excess of photons enhancing the effective photon temperature essentially only in the frequency range $65-90\,\MHz$ by a factor of $7/4$ with respect to the CMB temperature today, $T_{\rm CMB}$. The effective temperature is defined as
\be
T_{\rm eff} \equiv I/\mathcal{I}_{\rm BB} \,,
\ee
where
\be
I \equiv \int_{\rm E_{\rm min}}^{\rm E_{\rm max}} {\rm d}E \frac{\td F}{\td E} 
\ee
is the photon flux per solid angle in the energy range $(E_{\rm min},E_{\rm max}) = (65\,\MHz, 90\,\MHz)$, and 
\be
\mathcal{I}_{\rm BB} = \frac{E_{\rm max}^3-E_{\rm min}^3}{12\pi^2}
\ee
is the corresponding flux of black body radiation. In the following we will ascribe the net excess $T_{\rm eff} - T_{\rm CMB} = 3T_{\rm CMB}/4$ to photons directly emitted in DM decays or de-excitations.

In regard of this, interactions between photons and DM may allow the latter, or a fraction of it, to decay and contribute to the photon background. The relevant DM decay width in case of 2-body final states $X \to \gamma \tilde X$ or $X \to \gamma \gamma$ can be parametrised as 
\be
	\Gamma_{X} = E_\gamma^3/\Lambda^2 \,,
\ee
where $\Lambda$ is an effective mass scale and $E_\gamma$ is the energy of the emitted photons in the rest frame of $X$. For decays into two photons the photon energy is necessarily $E_\gamma = m_{X}/2$, while in de-excitation processes it is given by the mass splitting $E_\gamma \approx m_{X} - m_{\tilde X}$. Observing today an excess of photons in the band $E = 65-90 \, \MHz$  requires the photons to be produced between photon decoupling and today in the energy range
\be\label{range:Eg}
	3 \times 10^{-7}\,\eV< E_{\gamma} < 4 \times 10^{-4}\,\eV \,.
\ee

\begin{figure}[t]
\begin{center}
\includegraphics[width=.44\textwidth]{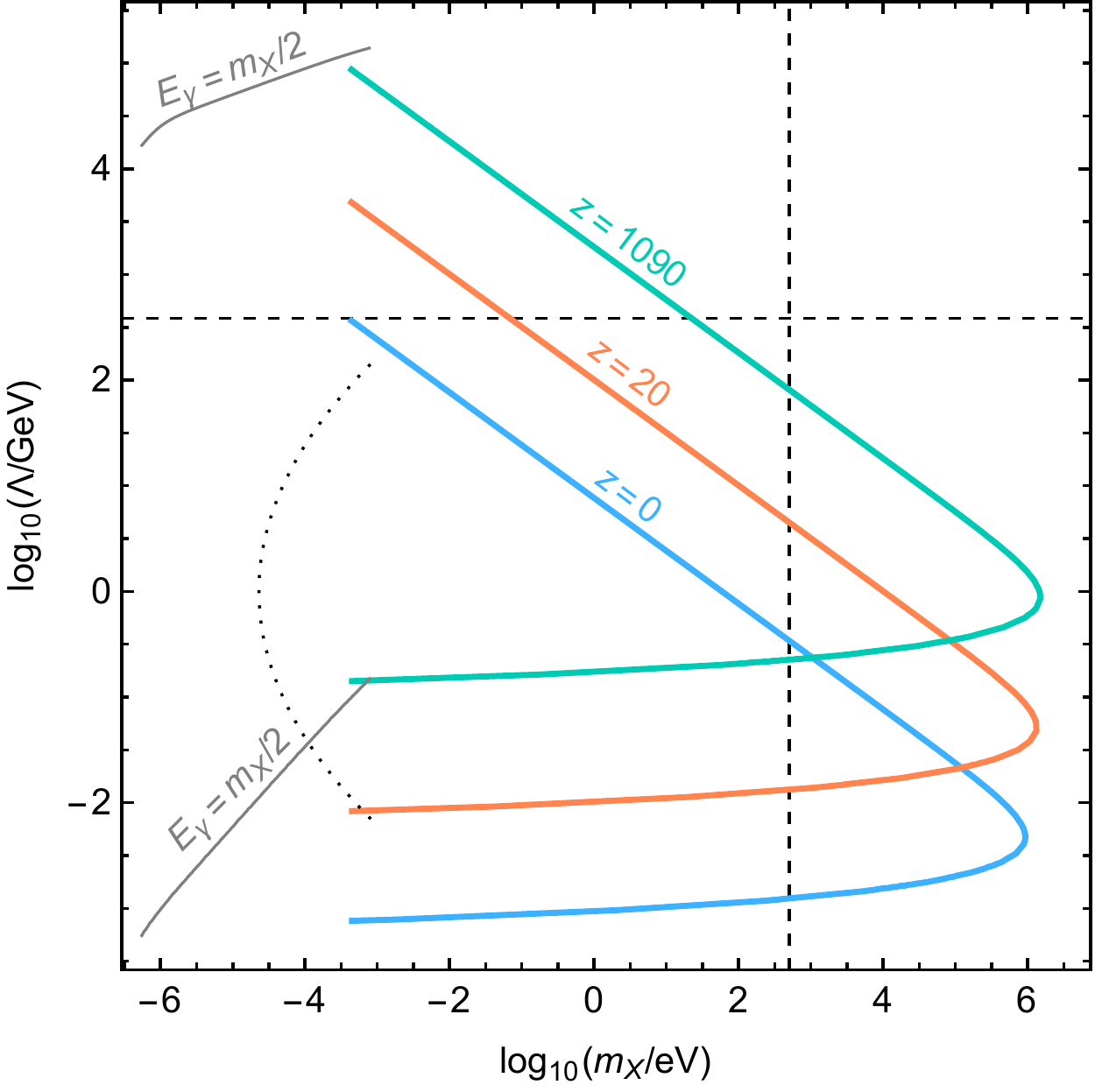}
\caption{The solid lines show the parameter values which produce the required photon excess assuming that all DM consists of $X$. The blue, red and green lines correspond to the case $X \to \gamma \tilde X$, with the labels showing the redshifts when the photons were emitted. The gray lines correspond instead to $X \to \gamma \gamma$. The mass bounds arise from Eq.~\eqref{range:Eg}, as described in the text. The region on the left of the vertical dashed line is excluded for fermionic DM by the Tremaine-Gunn bound. The area below the horizontal dashed line is excluded by the electroweak precision measurements in the case of exiting DM. In the region on the right of the dotted curve the lifetime of $X$ is shorter than the age of the Universe for the $X \to \gamma \gamma$ case.}
\label{fig:constr}
\end{center}
\end{figure}

The resulting differential flux of photons with energy $E$ is then given by~\cite{Masso:1999wj}
\be\label{dF/dE}
	\frac{\td F}{\td E}
	= \frac{A}{4\pi} n_{X,0} \Gamma_{X} \left.\frac{ e^{-\Gamma_{X}t(z)}} {H(z)}  \right|_{1+z = E_{\gamma}/E} \,,
\ee
where $A$ is the number of photons emitted per process and $H(z)$ the Hubble constant at a redshift $z$ corresponding to a time $t(z)$ such that $t(0)$ is the age of the Universe. We indicate with $n_{X,0}$ the present $X$ abundance if $X$ were not to decay. By requiring that these processes produce an effective temperature excess of $3T_{\rm CMB}/4$, we constrain the parameters of the scenario as shown in Fig.~\ref{fig:constr}. 

For two photon decays $X \to \gamma \gamma$ the allowed mass range $5 \times 10^{-7}\,\eV< m_X < 8\times 10^{-4}\,\eV$  is dictated by Eq.~\eqref{range:Eg}, while for $X\to \gamma \tilde X$ the lower bound on $E_{\gamma}$ implies $m_X > 4 \times 10^{-4}\,\eV$. The existence of solutions characterized by different values of $\Lambda$ for a given $m_X$ can be understood as follows: In the upper parts of the curves $\Gamma_X t(z)\ll 1$, so the exponential suppression is negligible and $m_X$ has to increase for increasing $\Gamma_X$ (corresponding to decreasing $\Lambda$) in order to ensure the required photon flux. Conversely on the lower parts $\Gamma_X t(z)\gg 1$, hence $m_X$ has to decrease for increasing $\Gamma_X$.

As indicated by the plot, the scale $\Lambda$ is bounded to be 
\be \label{range:Lambda}
\Lambda < 2\,\TeV \sqrt{\frac{A f_X}{m_X/\eV}},
\ee 
predicting the emergence of further new physics at $\Lambda\lsim 100\,\TeV$. Likewise, the DM mass cannot be arbitrarily large because of the number density of photons required for the excess. The maximal DM mass can be estimated by requiring that all DM particles emit one photon at a time, yielding $m\lsim 3\,\MeV$.

The constraints on $E_\gamma$, $\Lambda$ and $m_X$ depend on the specifics of the adopted DM model, which we discuss in the following.

\subsubsection{Scalar DM}

In case of spin 0 DM particles such as axions~\cite{Peccei:1977hh,Weinberg:1977ma} or ALPs~\cite{Marsh:2015xka,Ringwald:2012hr}, the decay to photons is described by the effective operators
\be\label{int:scalar}
	\frac{1}{4} g_{V} X F^{\mu\nu}F_{\mu\nu} + \frac{1}{4} g_{A} X F^{\mu\nu}\tilde F_{\mu\nu} \,.
\ee
The corresponding decay width is 
\be
	\Gamma_X =  \frac{E_\gamma^3}{8\pi}(g_{V}^2 + g_{A}^2) 
\ee
with $E_\gamma = m_{X}/2$. In the mass range indicated by Eq.~\eqref{range:Eg}, the coupling between axions and photons is severely constrained by helioscope experiments: $g_{A} \ll 10^{-10}\, \GeV^{-1}$~\cite{Patrignani:2016xqp, Inoue:2008zp}. This constraint applies also to the scalar coupling $g_{V}$~\cite{Masso:1995tw}, implying $\sqrt{8\pi/(g_{V}^2 + g_{A}^2)} = \Lambda \gtrsim 10^{10}\,\GeV$. To explain the EDGES anomaly we require, however, $\Lambda < 10^5\, \GeV$, hence the necessary radio background from axions or scalars with interactions given by \eqref{int:scalar} requires interactions several orders of magnitude stronger than axion searches allow.

\subsubsection{Spin-2 DM}

Another possible candidate of bosonic DM is spin-2 DM in the form of massive bigravitons ~\cite{Aoki:2013joa,Aoki:2016zgp,Babichev:2016bxi,Babichev:2016hir,Aoki:2017cnz,Marzola:2017lbt}. The interaction between photons and massive gravitons is given by
\be
	\frac{\alpha}{\mpl} X_{\mu\nu}T^{\mu\nu}_{\rm EM},
\ee
where $X_{\mu\nu}$ is the perturbation of the massive graviton around the vacuum, $\alpha$ is a dimensionless coupling constant and $T^{\mu\nu}_{\rm EM}$ is the electromagnetic stress-energy tensor. The decay width to photons is~\cite{Han:1998sg}
\be
	\Gamma_X = \frac{E_\gamma^3}{10\pi} \frac{\alpha^2}{\mpl^2 }
\ee
with $E_\gamma = m_{X}/2$ as before and $\mpl = 2.4 \times 10^{18} \, \GeV$ is the reduced Planck mass. On general grounds, the existence of such a massive graviton will however introduce a new universal short range force acting alike on every matter field. Short range gravitational experiments, which seek the effects of such a fifth force,  constrain $\alpha \lesssim 10^{-2}$ for $m_{X} \lesssim 10^{-3}\, \eV$~\cite{Kapner:2006si}, implying $\Lambda > 10^{21} \, \GeV$ and therefore clearly ruling out the spin-2 scenario.

\subsubsection{Excited DM}

Another exciting possibility is provided by DM candidate with electric or magnetic dipole interactions given by
\be \label{eq:this}
	-\frac{i}{2}F_{\mu\nu} \bar{X}  \sigma^{\mu\nu} (\mu_{X} + d_{X} \gamma^{5}) \tilde X + \mathrm{h.c.} \,,
\ee
where $\sigma^{\mu\nu} \equiv i[\gamma^{\mu},\gamma^{\nu}]/2$. This results in a decay rate into photons~\cite{Chang:2010en}
\be
	\Gamma_X = \frac{E_\gamma^3}{\pi}(\mu_{X} ^2 + d_{X}^2)
\ee
with $E_\gamma = m_{X} - m_{\tilde{X}}$. The constraints on the magnetic and electric dipole of DM from electroweak precision measurements force $\mu_{X}, d_{X} < 3 \times 10^{-3}e\, {\rm fm}$~\cite{Sigurdson:2004zp}. When $m_X\simeq m_{\tilde X}$, this constraint also applies to the dipole transition given by Eq.~\eqref{eq:this}, implying $\Lambda \gtrsim 400\, \GeV$. Although this bound is consistent with Eq.~\eqref{range:Lambda} for small DM masses, imposing the  Tremaine-Gunn bound $m_{X} f_X^{-1/4} \geq 0.5 \, \keV$~\cite{Tremaine:1979we,Boyarsky:2008ju} rules out the possibility of fermionic DM. Since the effective temperature scales as $f_X$, cases where only a fraction $f_X<1$ of DM is in the form of excited DM are also excluded.

Although the Tremaine-Gunn bound does not apply to bosonic DM candidates,  dipole interactions characterised by $\Lambda < 10^5\, \GeV$ can maintain $X$ in kinetic equilibrium down to temperatures below $1~ \MeV$~\cite{Gondolo:2016mrz}. Successful big bang nucleosynthesis requires $m_{X} > 1~\MeV$~\cite{Boehm:2013jpa}, which is in contradiction with Eq.~\eqref{range:Eg} if the bosons is in chemical equilibrium. As the constraint on the dipole moment of DM is expected to hold in order of magnitude regardless of the spin of DM, we can assume also for this case that $\Lambda \gtrsim 400\, \GeV$, resulting in a constraint on the DM mass: $m_{X} \leq 0.02 \, \keV$. Because a late kinetic decoupling results in a too warm DM candidate if $m_X \leq 2 \, \keV$~\cite{Yeche:2017upn}, we conclude that also the bosonic case is ruled out.

\subsection{Mini-clusters --  the cosmological  enhancement mechanism for ALP decays}
\label{sec:miniclusters}

While ALPs seem to be the most natural DM candidates to provide soft photons in the $65-90\,\MHz$ range, their decays as the solution to the 21 cm anomaly is clearly ruled out.
In order to find a viable solution to the anomaly, an enhancement mechanism for ALP decays must then exists. Indeed, non-linear effects in the evolution of the ALP field in the early Universe may lead to the formation of gravitationally bound, dense clumps of ALPs -- the so called  mini-clusters~\cite{Tkachev:1986tr,Hogan:1988mp,Kolb:1993zz}. Inside the mini-clusters, under some conditions, parametric instability~\cite{Tkachev:1986tr} may result in a powerful coherent burst of radiation that may have
already been observed as Fast Radio Bursts~\cite{Tkachev:2014dpa}. Interestingly, ALP mini-clusters also provide the cosmological mechanism that is needed for an ALP explanation of the 21 cm anomaly to be viable option.

In this regard, other cosmological observables bound the mass of oscillating coherent DM fields. Large scale structure formation sets a lower bound $m > 10^{-24}$ eV corresponding the inverse size of a dwarf galaxy on the axion mass \cite{Hlozek:2014lca,Hlozek:2016lzm}. Black hole superradiance, instead,  excludes the mass range $6 \times 10^{-13}~\mathrm{eV} > m > 2 \times 10^{-11} ~\mathrm{eV}$ \cite{Arvanitaki:2015iga,Arvanitaki:2016qwi}. In addition to that, pulsar timing arrays will soon test the lowest end of the mass range \cite{Khmelnitsky:2013lxt,Porayko:2014rfa}. Still, our considerations point to a much heavier DM candidate with mass in the $10^{-6}-10^{-3}$~eV range.

\section{Discussion and Conclusions} 
\label{sec:Conclusions}

In this paper we considered several DM scenarios that can modify \Eq{signal} in the direction indicated by the EDGES 21 cm measurement.
We pointed out robust arguments that disfavour the proposed solutions that rely on cooling of hydrogen gas via DM interactions. While no definitive conclusions
can be drawn by using only simplified proposals, we tend to agree with the conclusions in the literature and find scenarios based on this mechanism  severely constrained.

Motivated by this result we studied DM candidates that can directly modify the background soft photon temperature in \Eq{signal}. We pointed out that the DM processes that produce also a hard photon spectrum in addition to the soft component needed to explain the anomaly, are also generally constrained. This leads us to conclude  that decaying very light photophilic DM is needed to explain the 21 cm signal. 
We then studied axion and generic ALP decays, massive spin-2 DM decays, and fermionic and bosonic excited DM solutions to the 21 cm anomaly.
We found that at perturbative level, just considering decays of individual particles, all these solutions are ruled out by the interplay of the required lifetimes and dedicated experimental bounds. 

However, cosmological enhancement mechanisms that could rescue some scenarios exist at the non-perturbative level. In ALP or oscillating spin-2 DM mini-clusters, parametric instabilities may enhance the DM conversion into photons, leading to the explosions of the mini-clusters itself and providing the required soft photon emission. 
As a result of this work,  the axion, ALP or oscillating spin-2 DM mini-clusters augmented with a parametric instability mechanism constitute the only viable explanation to the EDGES 21 cm anomaly.

\medskip

\noindent{\bf Acknowledgements}  This work was supported by the grants IUT23-6, IUT26-2, PUT808, PUT799, and by EU through the ERDF CoE program grant TK133 and by the Estonian Research Council via the Mobilitas Plus grant MOBTT5.

\vspace{-5mm}
\bibliography{21cm}

\end{document}